\documentclass[a4paper,twoside]{article}
\newcommand{\affil}[1]{$^{\rm #1}$}
\usepackage[authoryear]{natbib}
\bibpunct{(}{)}{;}{a}{}{,}
\usepackage{graphicx}
\usepackage{amssymb}
\usepackage[breaklinks]{hyperref}
\usepackage{breakurl}

\date{} 
\def\HI{H{\sc i} }
\def\HII{H{\sc ii} }

\title{\large\bf\center Very High Angular Resolution Science with the \\Square Kilometre Array}
\author{\parbox{\textwidth}{\flushleft
\vspace{-0.5cm}
{\it L.E.H. Godfrey\affil{A,H}, H. Bignall\affil{A}, S. Tingay\affil{A}, L. Harvey-Smith\affil{B}, M. Kramer\affil{C,D}, S. Burke-Spolaor\affil{B,E}, J.C.A. Miller-Jones\affil{A}, M. Johnston-Hollitt\affil{F}, R. Ekers\affil{B,A}, \\ S. Gulyaev\affil{G}}\\
\vspace{0.4cm}
{\small \affil{A}\,International Centre for Radio Astronomy Research, Curtin University, GPO Box U1987, Perth, WA 6845, Australia}\\
{\small \affil{B}\,CSIRO Astronomy and Space Science, Australia Telescope National Facility, PO Box 76, Epping, NSW 2121, Australia}\\
{\small \affil{C}\,Max-Planck-Institut fur Radioastronomie, Auf dem Hugel 69, D-53121 Bonn, Germany}\\
{\small \affil{D}\,Jodrell Bank Centre for Astrophysics, University of Manchester, Manchester M13 9PL}\\
{\small \affil{E}\,Jet Propulsion Laboratory, California Institute of Technology, Pasadena, CA 91109}\\
{\small \affil{F}\,School of Chemical and Physical Sciences, Victoria University of Wellington, PO Box 600, Wellington, 6140, New Zealand}\\
{\small \affil{G}\,Institute for Radio Astronomy and Space Research, Auckland University of Technology, Auckland, New Zealand}\\
{\small \affil{H}\,Corresponding author. Email: L.Godfrey@curtin.edu.au}}}
\begin{document}
\twocolumn[
\begin{changemargin}{.8cm}{.5cm}
\begin{minipage}{.9\textwidth}
\vspace{-1cm}
\maketitle
%

\vspace{-0.85cm}
\begin{center}
{{\it Accepted for publication in PASA: November 24, 2011}}
\end{center}
\vspace{-0.25cm}

\small{\bf Abstract:}

Preliminary specifications for the Square Kilometre Array (SKA) call for 25$\%$ of the total collecting area of the dish array to be located at distances greater than 180 km from the core, with a maximum baseline of at least 3000 km. The array will provide angular resolution $\theta \lesssim$ 40 -- 2 mas at 0.5 -- 10 GHz with image sensitivity reaching $\lesssim 50$~nJy/beam in an 8 hour integration with 500~MHz bandwidth. Given these specifications, the high angular resolution component of the SKA will be capable of detecting brightness temperatures $\lesssim$ 200 K with milliarcsecond-scale angular resolution. The aim of this article is to bring together in one place a discussion of the broad range of new and important high angular resolution science that will be enabled by the SKA, and in doing so, address the merits of long baselines as part of the SKA.  We highlight the fact that high angular resolution requiring baselines greater than 1000 km provides a rich science case with projects from many areas of astrophysics, including important contributions to key SKA science.

\medskip{\bf Keywords:} telescopes

\medskip
\medskip
\end{minipage}
\end{changemargin}
]
\small

\section{Introduction}

The future of radio astronomy at cm-wavelengths lies with the Square Kilometre Array (SKA) --- a radio interferometer currently in the design stages, that will have a total collecting area in the order of one square kilometre \citep[see e.\,g.][]{schilizzi07}. The current design stipulates that 25$\%$ of the total collecting area will reside in a number of remote array stations at distances of between 180 km to 3000 km from the centre of the array. It is envisaged that each of these remote stations will comprise several dishes with single pixel receivers operating in the approximate frequency range 0.5 - 10 GHz \citep[e.\,g.][]{schilizzi07}. Such an instrument will provide very high sensitivity at angular resolutions ranging from several arcseconds to one milliarcsecond. 

In the following sections, we highlight the fact that high angular resolution requiring baselines greater than 1000 km provides a rich science case with projects from many areas of astrophysics, including important contributions to key SKA science. This was presented in SKA Memo 135 \citep{godfrey11}, and the following represents a subset of the science discussed in that work. 

We note that the SKA remote stations will not include the sparse or dense aperture array technologies proposed for the SKA core, nor will they involve phased array feeds on the dishes. For the remote stations, only the standard technologies --- dishes with single pixel feeds, will be involved. Much of the proposed high angular resolution science does not require access to very wide fields of view or very low frequencies ($< 500$ MHz). Therefore, dishes with single pixel receivers operating in the approximate frequency range 0.5 --- 10 GHz are adequate for the vast majority of proposed high angular resolution science. 

The science case will continue to develop as the SKA design proceeds over the coming years, as part of the Pre-construction phase Project Execution Plan or PEP \citep{schilizzi11}. 

In \S \ref{sec:science} we compile a list of science drivers for the high angular resolution component of the SKA. The science cases are extracted largely from the book ``Science with the Square Kilometre Array" \citep{carilli04}, and the Design Reference Mission for SKA-mid and SKA-lo \citep{drm}, and we include more recent advances and additional science cases. In \S \ref{sec:conclusions} we present the conclusions and closing remarks resulting from this work.

\section{High Angular Resolution \\Science Case} \label{sec:science}

\subsection{Strong Field Tests of Gravity} \label{sec:pulsar_BH_distances}

One of the major science goals of the SKA is to test relativistic theories of gravity in the strong field regime via  precision timing of pulsars. This will be achieved by (1) timing relativistic binary systems --- e.\,g.\ pulsar-neutron star binaries and any pulsar-black hole binaries discovered in the future, including pulsars in orbit around the Galactic Centre; and (2) monitoring an array of millisecond pulsars (a pulsar timing array) to detect gravitational waves with nanoHertz frequencies \citep{kramer04, cordes04}. This science goal is discussed in the Design Reference Mission \citep{drm} chapters 16 and 17. So far the discussions of the strong field gravity tests using pulsars have concentrated on the critical precision timing information, but the importance of high angular resolution has not been emphasised. Here, we highlight the importance of high angular resolution to achieve the aims of this important science goal. 

Approximately 100 compact relativistic binaries are expected in the SKA Galactic pulsar census \citep{smits09}, of which some fraction ($\gtrsim 5 - 25$) are expected to be in stellar-mass black hole binary systems \citep{lipunov05}. The likelihood of dynamic interactions in globular clusters means that the chances of finding exotic binaries such as millisecond pulsar-BH systems is enhanced in these environments \citep[e.\,g.][]{sigurdsson03}. However, the most common BH-pulsar binary system are likely to be normal rather than recycled (millisecond) pulsars \citep[see][]{pfahl05, lipunov05}. 

In this section dealing with strong field tests of gravity, we focus on relativistic binaries in which the pulsar companion is a stellar-mass black hole, neutron star, or white dwarf. The discovery of a pulsar in a sufficiently compact orbit around the supermassive black hole at the Galactic centre (GC) would also enable tests of relativistic gravity that are complementary to those enabled by pulsars in compact orbits around stellar-mass black holes. The prospects of probing the space-time of the supermassive black hole at the Galactic centre via pulsar timing measurements are discussed in detail by \citet{liu11}.  For GC pulsars, an orbital period $\lesssim$ 0.3 yrs would be required to ensure that perturbations caused by the mass distribution around Sgr A* are negligible \citep{liu11}. Furthermore, frequencies $\gtrsim$ 15 GHz would be required to optimise the timing precision \citep{liu11} which is strongly affected by pulse-broadening caused by the extreme interstellar scattering at the GC. 

\subsubsection{Accurate pulsar distances are \\essential}

Precise measurements of the proper motion and distance to each of the relativistic binaries detected in the pulsar census are essential for these systems to be used as laboratories for testing theories of gravity. Accurate distance and proper motion measurements are required in order to correct for the acceleration terms that affect the spin and orbital period derivatives. The latter parameter is of particular relevance for testing alternative theories of gravity \citep{cordes04, stairs10, kramer10} and potentially detecting, or at least constraining, extra spatial dimensions \citep{simonetti11}. Let $P_b$ be the binary period, $\dot{P_b}$ the corresponding time derivative, c the speed of light, d the distance and $\mu$ the proper motion of the system. The so-called Shklovskii-effect \citep{shklovskii70} contributes to the observed period derivative an amount
\begin{equation}
\frac{\dot{P}_b}{P_b} = \frac{\mu^2 d}{c}
\end{equation}
This effect, if not precisely accounted for, limits the precision with which theories of gravity may be tested in relativistic binary pulsars. In some cases, the magnitude of the Shklovskii-effect can be comparable to, or greater than the intrinsic orbital period derivative due to gravitational radiation \citep[see e.\,g.][]{bell96}. A similar effect arises due to the differential acceleration of the solar system and the pulsar in the gravitational potential of the Galaxy \citep{damour91}. The determination of this Galactic acceleration term requires precise knowledge of the pulsar's spatial position, as well as the Galactocentric radius ($R_0$) and speed of the solar system ($v_0$). To underline the importance of precise distance measurements, it is worth noting that the tests of relativistic gravity in the Hulse-Taylor binary system B1913+16, which currently provides one of the most precise constraints of this kind, are limited by the uncertainty in the distance, which has been determined using the pulsar's dispersion measure to a precision of $\sim 30\%$ \citep{weisberg08}. 

As noted above, the Galactic constants $R_0$ and $v_0$ are of fundamental importance in correcting for the acceleration terms that impact the observed binary period derivative. The high angular resolution component of the SKA could provide a measurement of $R_0$ with $1\%$ precision via parallax measurements of Sgr A* \citep{fomalont04}. 

\begin{figure*}[ht]
\centering
\includegraphics[scale=0.3]{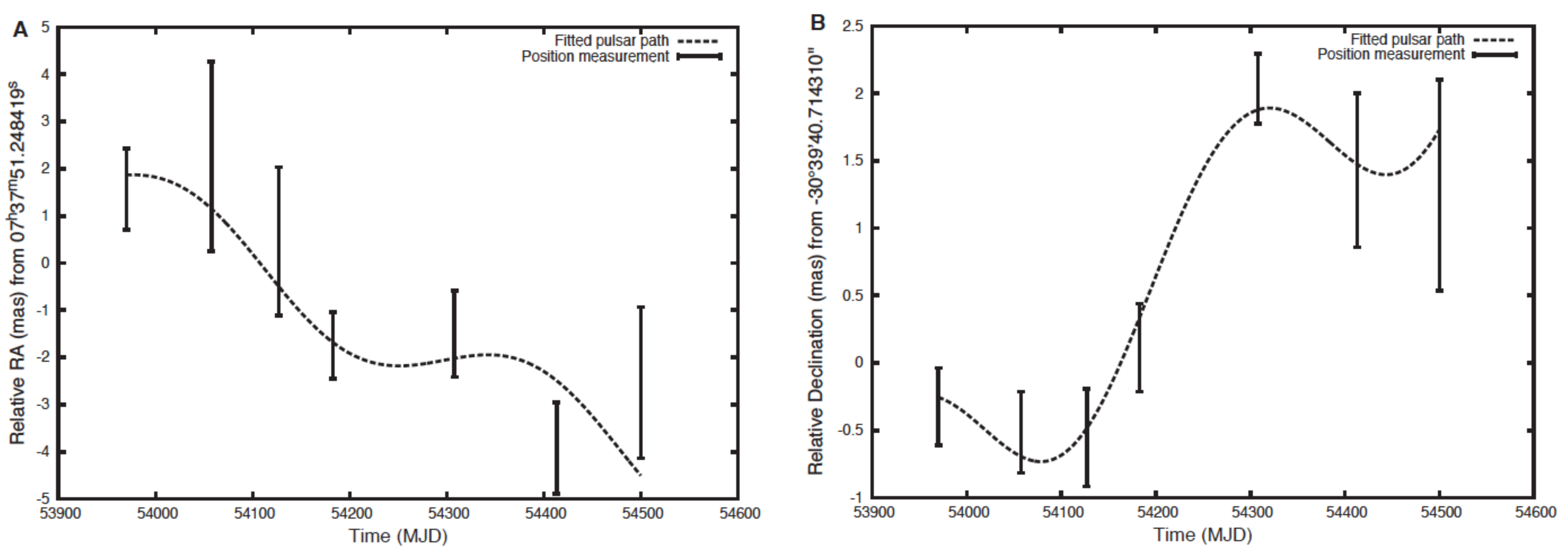}
\caption{ Motion of PSR J0737-3039A/B plotted against time. Trigonometric parallax measurements for this relativistic binary pulsar system revealed that the distance was more than a factor of 2 greater than previous distance estimates based on dispersion measure and timing parallax measurements. The precise interferometric distance and proper motion measurements combined with a decade of additional timing data will enable tests of GR at the 0.01$\%$ level using the orbital period derivative of this system \citep[][]{deller09}. Figure reproduced from Deller, Bailes \& Tingay, 2009, Science, Vol. 323, pg.\ 1327, with permission from The American Association for the Advancement of Science.} \label{fig:deller_pulsar_parallax}
\end{figure*}

\subsubsection{Trigonometric parallax measurements are required to maximise the science return}

Pulsar distances, in some cases, may be determined by timing measurements alone via the method of timing parallax. The orbital motion of the Earth causes a 6 monthly variation in the pulse arrival times due to the curvature of the wavefront, and consequent periodic change in the path length from the pulsar to Earth. 
The amplitude of this timing parallax signature is very small: ~$\Delta t_{\pi} \approx  1.2 \rm{\mu s} \times \cos \beta ~ d_{\rm kpc}^{-1}$,  ~~where $\beta$ is the ecliptic latitude, and $d_{\rm kpc}$ is the pulsar distance in kpc \citep{ryba91}. Therefore, accurate timing parallax measurements are limited to a subset of pulsars with very high timing precision; that is, millisecond pulsars with stable timing characteristics, and preferably low ecliptic latitude \citep{smits11}. In contrast, the ability to determine trigonometric (imaging) parallax (Figure \ref{fig:deller_pulsar_parallax}) depends only on the flux density and distance of the source, and is therefore applicable to a much wider range of systems. 

\begin{figure*}[ht]
\centering
\includegraphics[scale=0.3]{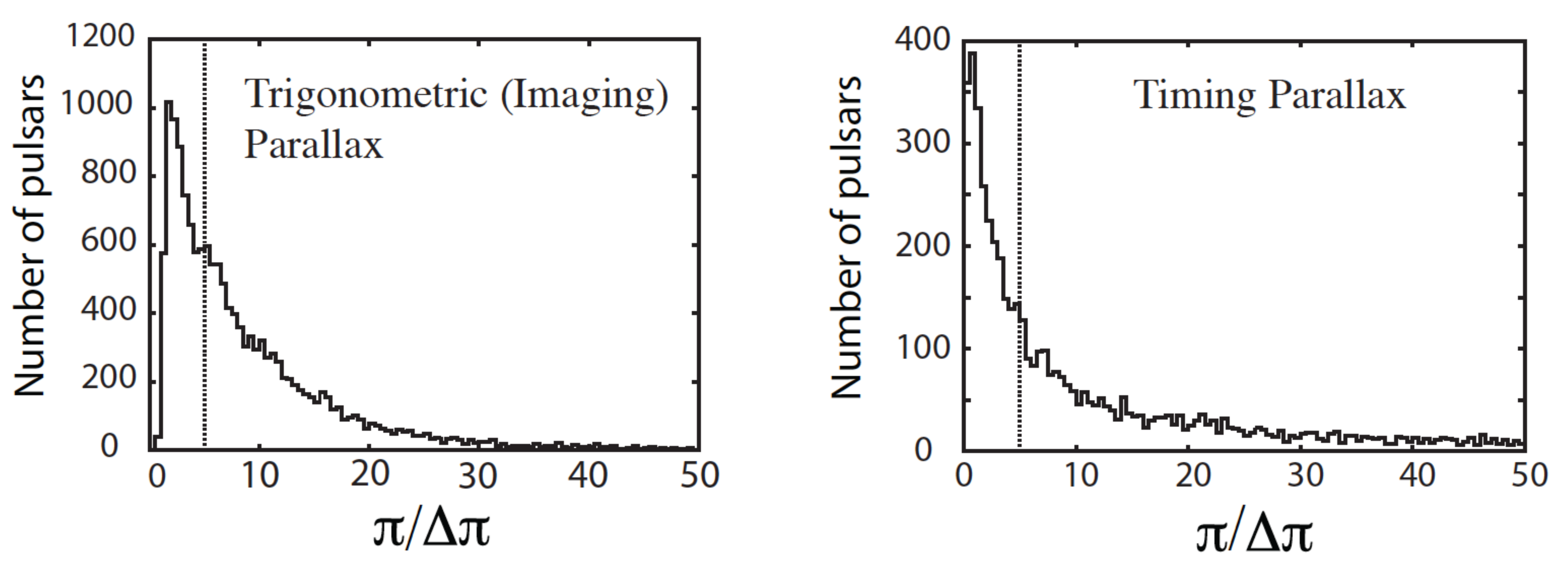}
\caption{From \citet{smits11}. Comparison between imaging and timing parallax histograms for the quantity $\pi / \Delta \pi$, where $\pi$ is the parallax and $\Delta \pi$ is the estimated error in the parallax for a simulated Galactic pulsar population. The vertical dotted lines mark the $\pi / \Delta \pi$ = 5 cutoff (20$\%$ error). \textbf{(Left)} Histogram of $\pi / \Delta \pi$ for trigonometric parallax measurements with the high angular resolution component of the SKA. The SKA can potentially measure the trigonometric (imaging) parallaxes for $\sim$9000 pulsars with an error of 20$\%$ or better. This includes pulsars up to a distance of 13~kpc. \textbf{(Right)} Histogram of $\pi / \Delta \pi$ for the timing parallax measurements of 6000 millisecond pulsars detected in the simulated SKA galactic pulsar census. Timing parallax measurements are limited to millisecond pulsars with very high timing precision, and therefore will not be possible for many pulsars detected in the Galactic pulsar census. The SKA can potentially measure timing parallax distances for about 3600 millisecond pulsars out to 9 kpc, with an error of 20$\%$ or better. Credit: Smits et al., A\&A, 528, A108, 2011, pages 5 \& 6, reproduced with permission $\copyright$  ESO.\label{fig:smits11_histograms}}
\end{figure*}

\citet{smits11} simulated and compared the accuracy of trigonometric parallax measurements with various methods of timing parallax distance determination, and concluded that both timing parallax and trigonometric parallax capabilities will be required to enable precision tests of gravity in the strong field regime. The results of the simulations (Figure \ref{fig:smits11_histograms}) suggest that the SKA can potentially measure the trigonometric parallax distances for $\sim$9000 pulsars up to a distance of 13 kpc with an error of 20$\%$ or better, and timing parallax distances for only about 3600 millisecond pulsars out to 9 kpc, with an error of 20$\%$ or better. 

It is highly likely that some of the most interesting relativistic binary systems will not provide sufficient timing precision to allow accurate timing parallax distance determination, but could still provide excellent tests for relativistic theories of gravity. This is possible because, despite the limited timing precision, accurate measurement of long term secular trends such as the orbital period derivative, $\dot{P}_b$, can still be achieved, given a long enough time. For example, the measured uncertainty in $\dot{P}_b$ decreases approximately as T$^{-2.5}$, where T is the total time span of data for the system \citep{damour92}. 

A good example of this is the pulsar-white dwarf relativistic binary system, J1141-6545. Owing to the asymmetry in self-gravitation between the pulsar and white dwarf companion, this system provides a unique laboratory for testing alternative theories of gravity \citep{bhat08}. However, the young pulsar in this system exhibits significant ``timing noise" which limits the timing precision \citep{bailes05}. Despite the timing noise, J1141-6545 is likely to provide some of the most stringent tests of alternative theories of gravity: already four post-Keplerian parameters have been measured, and the orbital period derivative for this system is expected to be determined to better than $2\%$ by 2012, at which point uncertainty in the kinematic Doppler term, or Shklovskii-effect (the term involving the pulsar distance and proper motion) will dominate the errors \citep{bhat08}. With this example in mind, it should be noted that many of the pulsar-black hole binaries are likely to be normal pulsars (and probably young pulsars like J1141-6545, due to evolution of the systems), rather than recycled (millisecond) pulsars \citep[see][]{pfahl05, lipunov05}. This suggests that trigonometric (imaging) parallax measurements will be required to determine accurate distances for a large fraction of pulsar-black hole binaries. 

\subsubsection{Why is the SKA required?}

The high sensitivity of the long baseline SKA is required not only to detect weak and distant pulsars, but also to provide a high density of calibrator sources surrounding the pulsars that will enable multi-view in-beam calibration, and therefore high precision astrometry \citep{rioja09, fomalont04}. Owing to its high sensitivity, the long baseline component of the SKA will be able to perform multi-view in-beam calibration using several compact, closely spaced calibrator sources, the closest of which will be in the order of several arcminutes from the target \citep[see][]{godfrey11}. This technique will provide extremely accurate phase calibration at the position of the target, and provide astrometric precision of order 15$\mu$as at 1.4 GHz \citep{fomalont04}. Observations at frequencies below $\sim$ 5 GHz are affected by ionospheric refraction, but the ionospheric effects may be calibrated out using a wide bandwidth \citep{brisken00}. Only with the substantial improvement in sensitivity provided by the SKA will high precision astrometry on weak pulsars (and other weak sources) be possible. 

\subsubsection{Benefits of high angular resolution to the pulsar timing array}

High angular resolution could also be important in establishing the pulsar timing array (PTA) \citep{smits11}. Accurate astrometric information reduces the amount of observing time required to obtain a coherent timing solution by breaking the degeneracies between position uncertainty and pulsar spin-down \citep{smits11}. In the absence of accurate positional information, this can take 12 months or more. Therefore, the high angular resolution component of the SKA will assist the selection of stable millisecond pulsars to be included in the pulsar timing array. 

Further, the PTA may detect the gravitational wave signal from individual nearby binary black holes. In that case, precise distances to the pulsars in the PTA are required to enable a precise measurement for the gravitational wave source location \citep{lee11}. 

Lastly, the high angular resolution component of the SKA will compile a significant sample of SMBH binaries (see \S \ref{sec:SBBH}). The identification of a large sample of SMBH binaries would enable statistical studies of the inspiral rates in various phases of the binary evolution. The inspiral rates, and the possible existence of a ``stalling radius" are important factors in the interpretation of the gravitational wave background that will be investigated with the pulsar timing array \citep{jaffe03}.

\subsection{Modelling the Large-Scale Galactic Magnetic Field using Pulsars} \label{sec:magnetic_fields}

Wavelet tomography using a grid of thousands of pulsars with known rotation measures (RMs), dispersion measures (DMs) and distances will provide the best possible map of the Galactic magnetic field and electron density on large ($\gtrsim 100$~pc) scales \citep{stepanov02, noutsos09, gaensler04, beck04, gaensler06}. So far the discussions of mapping the Milky Way magnetic field using pulsars have concentrated on the ability to search for and identify many thousands of pulsars, but the importance of high angular resolution has not been emphasised. Here, we highlight the importance of high angular resolution to achieve the aims of this important science goal.

The DM and RM for a grid of thousands of pulsars will be obtained via the SKA Galactic pulsar census. The final ingredient to enable accurate tomographic models of the large scale Galactic magnetic field -- accurate distance estimates to each of the pulsars -- will require trigonometric parallax measurements to thousands of pulsars. Currently, distance estimates to pulsars are most commonly obtained via the pulsar's dispersion measure combined with the galactic electron density model. Distance estimates using this method are typically uncertain by tens of percent, and can be in error relative to accurate parallax measurements by more than a factor of 2, due to the large uncertainty in the electron density model \citep{deller09b}. Precise pulsar distances will require either parallax distance measurements, or an improved electron density model, which itself will require parallax distance measurements to a large sample of pulsars \citep{cordes04}. Therefore, precision astrometry is a requirement for the SKA to enable the best possible model of the large scale Galactic magnetic field. Mapping the magnetic field of the Milky Way provides an excellent opportunity to address the issues surrounding the generation and preservation of galactic magnetic fields. The importance of understanding the large scale Galactic magnetic field configuration in the context of fundamental questions of astrophysics is discussed at length in e.\,g.\ \citet[][and refs. therein]{gaensler04, beck04}.

\subsection{Imaging protoplanetary disks at centimetre wavelengths}  \label{sec:ppd}

\begin{figure}[ht!]
\centering
\includegraphics[scale=0.28]{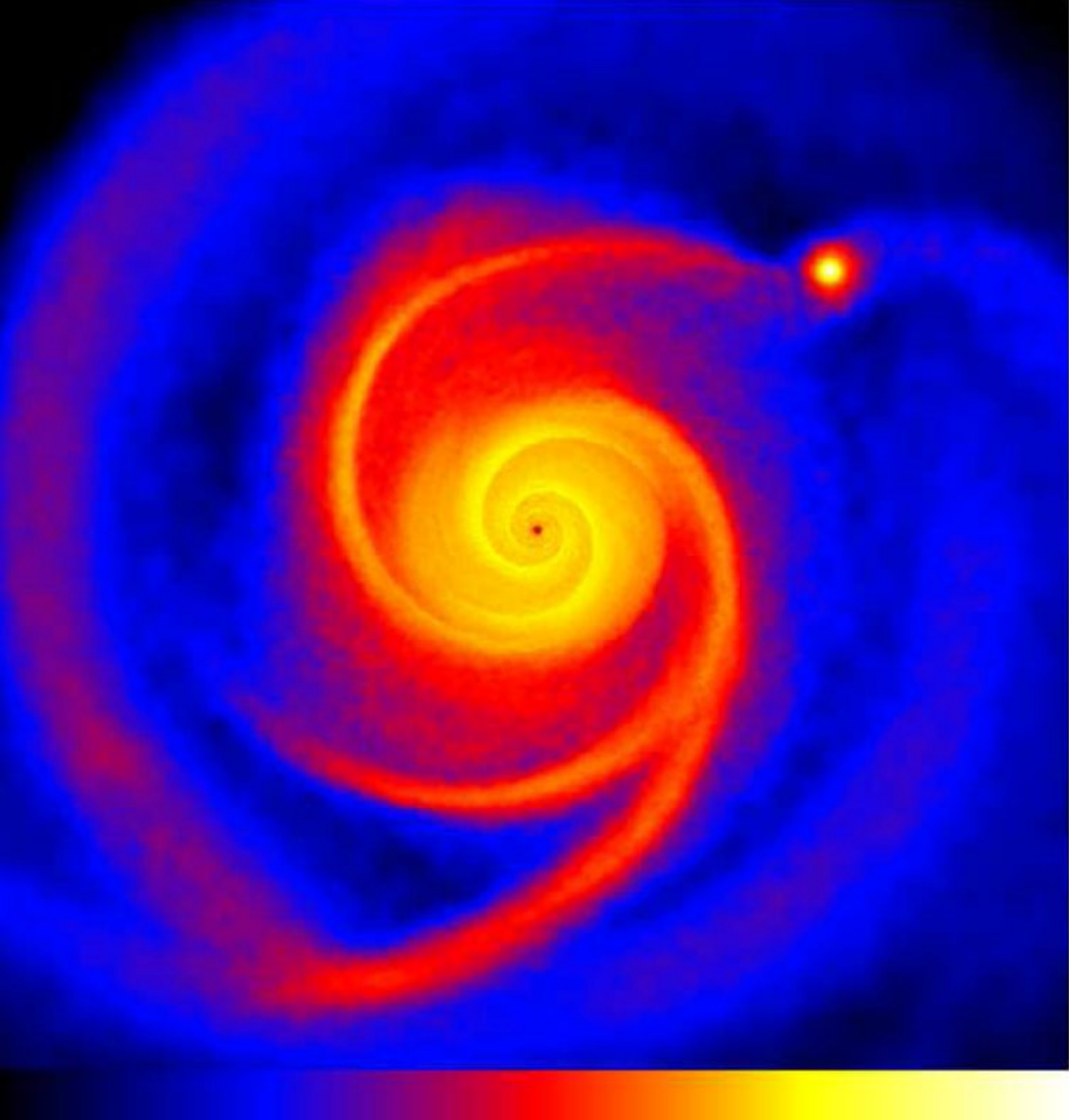}
\caption{Image of surface density structure in a protoplanetary disk from a smooth particle hydrodynamics simulation. This image shows the surface density structure of a 0.3 M$_\odot$ disk around a 0.5 M$_\odot$ star. A single dense clump has formed in the disk (upper right), at a radius of 75 AU and with a mass of $\sim$ 8 M$_{\rm Jupiter}$. Figure reproduced from Greaves et al., 2009, MNRAS: Letters, Vol.\ 391, pg.\ L76, with permission from John Wiley and Sons. \label{fig:ppd_simulation}}
\end{figure}

The scientific motivation for obtaining high angular resolution radio images of protoplanetary disks (Figure \ref{fig:ppd_simulation}) is three-fold. Firstly, it will enable imaging of various structures in the disk such as density waves and radial gaps formed by the interaction of the disk with a planetesimal \citep{wilner04}. Secondly, it will enable studies of the spatial dependence of spectral signatures relating to different grain properties in the disk \citep{greaves09}. Thirdly, imaging the \HI 21cm line emission will probe the kinematics and effects of photoevaporation in the disk surface layers \citep{kamp07}.

Grains in protoplanetary disks grow from sub-micron sizes up to mm sizes by sticking together in low-velocity collisions. Larger grains tend to shatter in collisions rather than sticking together. How, and under what conditions, do the mm-sized grains overcome this barrier to become pebble sized grains? This question is the subject of ongoing debate, and is a question that may be addressed with the high angular resolution component of the SKA. Dust particles emit inefficiently at wavelengths larger than their size, and therefore emission at cm-wavelengths provides evidence for pebble sized grains, which in turn provides evidence for significant progress towards planet formation. The high angular resolution component of the SKA will address the following questions: Where does the growth of decimetre-sized grains occur within the disk? Are the grains clumping into protoplanets? In what environments do these large grains occur (stellar age, spectral type, etc.)? Such information will benefit our understanding of planet formation and improve models of protoplanetary disks \citep{wilner04, wilner05, greaves09, natta07}. The reader is referred to \citet{greaves09} for a more detailed discussion of the science case for imaging protoplanetary disks at cm-wavelengths.

Imaging protoplanetary disks with the SKA was initially proposed for frequencies in the range 20 -- 35 GHz \citep{wilner04}. However, studies of protoplanetary disks can be carried out in the frequency range $\lesssim$ 10 GHz \citep{hoare09, greaves10, greaves09}. It is expected that the SKA will be able to image in detail the distribution of large dust particles in the disks around hundreds of nearby young stars at $\nu \lesssim 10$ GHz \citep{wilner05}. Initial estimates of the technical requirements indicate the need for very high sensitivity ($\sim$ 100 nJy/beam) on long ($\sim$ 1000 km) baselines \citep{greaves10}. This would enable $\sim$ 5 - 10 GHz observations of Earth analogues forming in southern star clusters  at $\sim$ 20 - 60 pc (the $\beta$ Pic, TW Hya, AB Dor, Tuc/Hor groups). The e-MERLIN Legacy Project ``PEBBLES" is aimed at studying the centimetre emission from pebble sized dust grains to show where and when planet-core growth is proceeding, and to identify accreting protoplanets. The initial results of the PEBBLES e-MERLIN survey will help to inform the scientific and technical requirements for this project with the SKA. 

\citet{kamp07} propose that mapping the \HI line in nearby systems will also be an important tool for studying circumstellar disks with the high angular resolution component of the SKA. Neutral Hydrogen 21cm line emission traces a layer near the disk surface that is directly exposed to soft UV irradiation from the parent star, but shielded from the ionising UV and X-ray ($h \nu > 13.6$ eV) radiation by the outer layer of the disk. High angular resolution SKA observations of 21cm line emission will probe the kinematics of protoplanetary disks, as well as the effects of irradiation and photoevaporation at the surface layer.

In addition to these primary scientific motivations, high angular resolution could potentially be used to pin-point the location of any extra-terrestrial intelligence (ETI) signals detected from planets orbiting relatively nearby stars \citep{morganti06}, by direct imaging and measuring the orbit of the planet. 

\subsection{Resolving AGN and Star \\ Formation in Galaxies}

At sub-mJy flux densities, the radio source counts at GHz frequencies are thought to be dominated by star-forming galaxies, as opposed to AGN which dominate source counts at higher flux densities \citep[e.\,g.][]{seymour08}. Without morphological information or a measurement of brightness temperature, it is generally not possible to determine, for a given galaxy, whether the observed radio flux is dominated by emission from a compact, nuclear starburst or an active galactic nucleus \citep{norris90}. The brightness temperature of a radio source indicates which process, AGN or star formation, dominates the radio emission: starbursts are typically limited to brightness temperatures of $T_b \lesssim 10^5$ K, and this clearly distinguishes them from the compact cores of AGN, which exhibit brightness temperatures $T_b >> 10^5$ K \citep{norris90, condon92}. Baselines longer than 3000 km are required to unambiguously distinguish AGN and star-formation in sources up to redshift $z = 7$ with flux densities down to at least 30 $\mu$Jy \citep[][chapter 2]{drm}. Discriminating between AGN and starburst galaxies will be possible in most cases based on the morphological information provided by high angular resolution images \citep[eg.][]{garrett99}. 

It is widely believed that AGN play an important role in the growth and evolution of galaxies. The interaction between the AGN and the surrounding medium may promote star formation at high redshift \citep[e.\,g.][]{klamer04, elbaz09} and/or suppress star formation at lower redshifts \citep[e.\,g.][]{croton06}. A powerful approach to addressing questions on the relationship between AGN activity, black hole growth, and galaxy evolution, will be deep, high-resolution imaging with the SKA to detect and distinguish between the first starburst galaxies and the first AGN jets, and to determine the frequency of occurrence of low luminosity AGN in different galaxy types \citep[][chapter 2]{drm}. This will enable a determination of the full range of SMBH masses and accretion rates and how these relate to galaxy histories.

This aspect of high angular resolution SKA science is discussed in detail in the SKA Design Reference Mission \citep[][chapter 2]{drm}. The goal will be to conduct a high angular resolution SKA survey to obtain a statistically significant sample of galaxies through which to explore the contribution and role of AGNs versus star formation in galaxy evolution. The high angular resolution SKA survey will be coordinated with other multi-wavelength surveys to maximise the scientific return, and an additional benefit will be in studying the cosmic evolution of AGN activity, which will address important questions relating to radio AGN, such as the lifetimes, duty-cycles, fuelling and triggering mechanisms. 

\subsection{The first generation of AGN jets}

The discovery of powerful distant quasars at $z \gtrsim 6$ indicates that supermassive black holes $>10^9 M_\odot$ existed at that time. This suggests that the first supermassive black holes formed before, or during, the epoch of reionisation. Indeed, it has been suggested that AGN jets may have played a key role in the formation of some of the first stars and galaxies in the universe, through jet-induced star formation \citep{klamer04, silk05, elbaz09, elbaz10}. 

\begin{figure}[h!]
\centering
\includegraphics[scale=1.03]{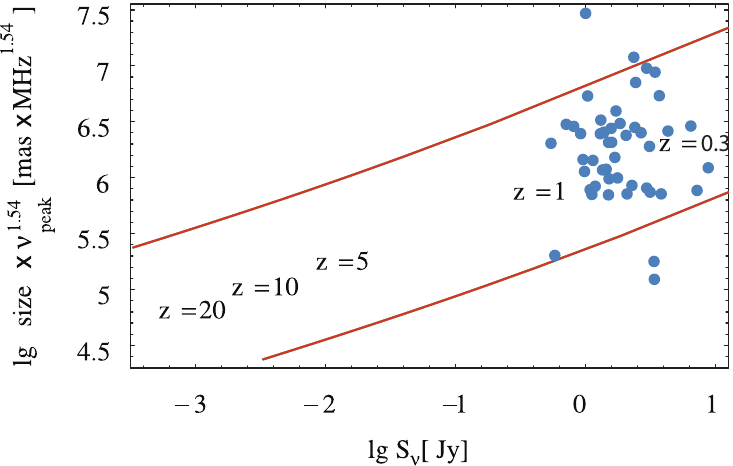}
\caption{Plot of a combination of the turn-over frequency and angular size (size $\times \nu_{\rm peak}^{1.54}$) versus the peak flux density
for a sample of GPS sources. Size, turn-over frequency, and flux density roughly form a
fundamental plane for GPS radio galaxies. Standard GPS sources found at $z \sim 1$ occupy the upper right of the plot. High redshift ``GPS-like" sources are expected to stand out from their low redshift counterparts, and occupy the lower left portion of the plot. See \citet{falcke04a} for details. Figure reproduced from New Astronomy Reviews, Vol.\ 48, Falcke et al., ``Compact radio cores: from the first black holes to the last", pg.\ 1169, Copyright 2004, with permission from Elsevier. \label{fig:first_SMBH}}
\end{figure}

\citet{falcke04a} suggest that the first generation of AGN jets produced by accreting supermassive black holes will be strongly confined by their dense
environment and appear as distant Gigahertz Peak Spectrum (GPS)-like sources --- that is, faint, compact sources with unusually low turn-over frequencies. The turn-over frequency, $\nu_{\rm peak}$ and linear size, $L$, of GPS and Compact Steep Spectrum (CSS) sources are found to follow an expression of the form 
\begin{equation}
\nu_{\rm peak} = 0.62 \left( \frac{L}{\rm kpc} \right)^{-0.65} \> \rm{GHz},
\end{equation}
which results from the basic properties of synchrotron self-absorption \citep{falcke04a}. Since the source size and turn-over frequency of GPS sources are correlated but angular size and frequency scale differently with redshift, the first AGN jets should stand out from their low redshift counterparts in the parameter space defined by angular size, turn-over frequency, and flux density (see Figure \ref{fig:first_SMBH}). 

\noindent \citet{falcke04a} suggest the following strategy for finding the first generation of AGN jets in the universe: 
\begin{itemize}
\item a shallow all-sky multi-frequency survey in the range 100 -- 600 MHz down to 0.1 mJy at arcsecond resolution;
\item identification of compact, highly peaked spectrum sources in that frequency range;
\item  identification of empty fields in the optical;
\item re-observation to exclude variable sources;
\item observations with long baselines and resolutions of $\sim$10 mas to determine sizes and to pick out the ultra-compact low-frequency peaked (ULP) sources;
\item spectroscopic confirmation of remaining candidates with \HI observations or by other means.
\end{itemize}
The stated goal of 10 mas resolution, at a frequency of 1.4~GHz, would require baseline lengths up to $\sim$ 4000~km.  \\

\subsubsection{Radio/CO Studies of high redshift AGN Jets}

\citet{klamer04} reviewed molecular gas observations for a sample of $z>3$ galaxies, and found that the gas and dust are often aligned with the radio emission. Based on these results, they proposed a scenario in which CO is formed at the sites of star formation that are triggered by relativistic jets, as is seen in some nearby sources (eg. Cen A, 3C40). High sensitivity, high angular resolution imaging of high redshift radio galaxies will be required to complement high redshift CO imaging with ALMA, in order to study the relationship between radio jets and early star formation. Resolution of order tens of mas will likely be required at low frequency ($\lesssim 1.4$~GHz) to map the radio structures in detail.

\subsection{Exploration of the Unknown}

The Exploration of the Unknown has been identified as an important guiding principle for the design of the SKA \citep{carilli04, wilkinson04}. This recognises the discovery potential provided by instruments that are capable of probing unexplored regions of parameter space. Whilst high angular resolutions are reached with existing radio telescopes, this domain has not been explored at the sensitivity of the SKA. The combination of high sensitivity and high angular resolution with the SKA will increase the observational phase space being searched, by opening up a large, unexplored region of the flux density--angular size plane. Observations at mas-scale resolution will, for the first time, be possible for thermal and non-thermal emission regions with brightness temperatures as low as hundreds of Kelvin. Current VLBI networks are, in general, limited to non-thermal sources with brightness temperatures $\gtrsim 10^6$~K. The combination of high sensitivity with a broad range of angular resolution up to mas-scales will provide greater discovery potential for the SKA. Furthermore, the ability to perform high angular resolution follow-up of transient radio sources will maximise the science return of transient searches, as discussed below.

\subsubsection{Transients}

High angular resolution will play an important role in localising,
identifying and understanding transient radio sources. Arcsecond
resolution may be sufficient to identify the host galaxies of
extragalactic fast transients, and follow-up spectroscopy of the host
galaxies would provide the redshifts. However, mas-scale resolution could
potentially localise transient sources on a much finer scale and help
to determine their nature. High angular resolution follow-up
observations of newly discovered classes of radio source would be of
great benefit to understanding the source physics. Resolving the source morphology and its evolution could
provide information on the energetics of the event and environment of
the source. For the slower transient sources (with time-scales of weeks or longer), high angular resolution
would enable measurement of the source proper motion which could discriminate between Galactic and extragalactic events.
This would be particularly important if sources were found to be unresolved with no optical counterparts.

Long baselines are also an excellent discriminant between RFI and
genuine astronomical events \citep{wayth11, thompson11}. A triggered
buffer \citep[e.\,g.][]{macquart10b} would allow
for off-line analysis of the transient sources, and would function as follows: 

\begin{itemize} 
\item Data from antennas on
long baselines would be stored for a couple of minutes in a rolling
buffer. 
\item A transient source detected within the long baseline field-of-view (effectively the 15m antenna primary beam) would
trigger the download of this buffer for post-processing. 
\item The station beams could then be formed in the direction of the transient source whose
location would be determined by the SKA core to within a few
arcseconds. A rolling buffer is not required for the antennas of the SKA core, since these antennas will have access to the whole field of view, and the standard output would enable the transient source position to be determined to within a few arcseconds. 
\end{itemize}
A pilot survey (V-FASTR) for VLBI detection of fast
transients using a triggered buffer is currently being implemented
on the VLBA \citep{wayth11}. The results of the V-FASTR survey
will inform the technical requirements for this experiment
with the SKA.

\subsection{Binary Supermassive Black Holes} \label{sec:SBBH}

\begin{figure*}[ht]
\centering
\includegraphics[scale=0.9]{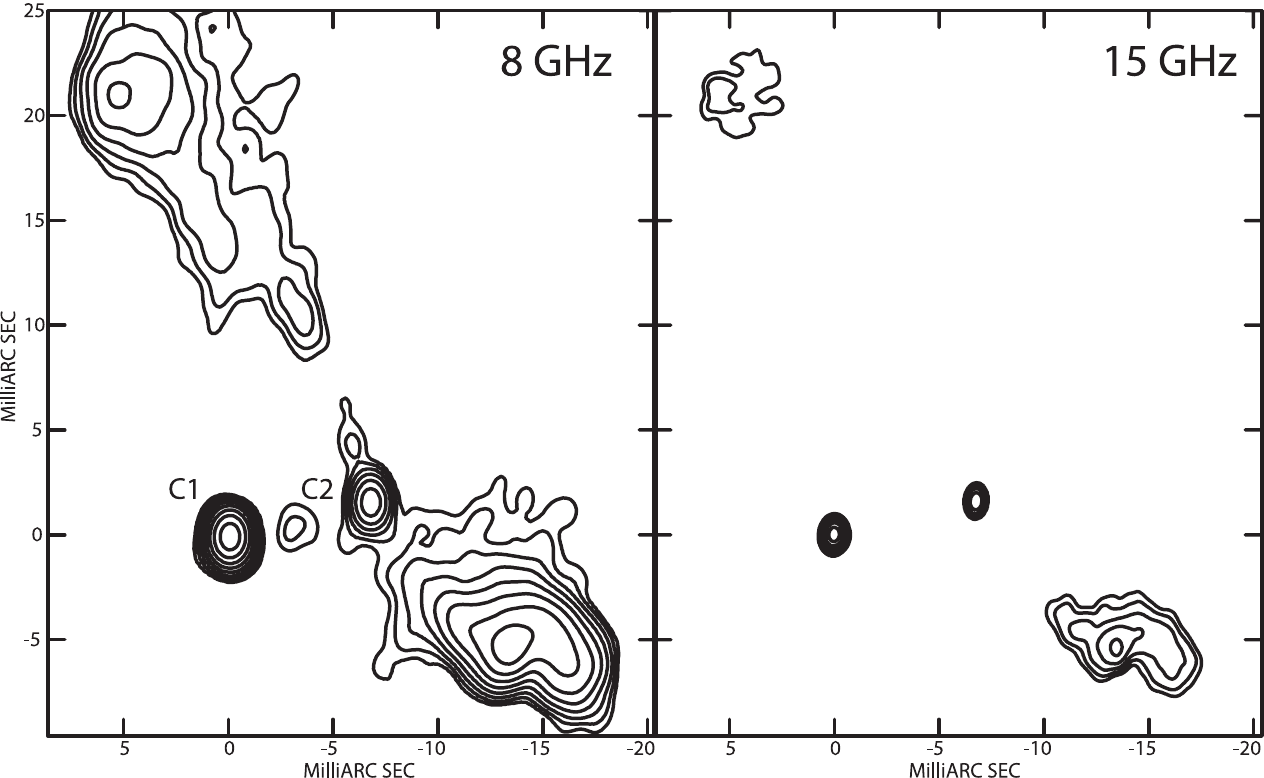}
\caption{VLBA images of the binary black hole system 0402+379 at 8 and 15 GHz. The pair of unresolved, flat spectrum radio cores are easily identified in this sequence of images. The projected separation between the two black holes is 7.3~pc. Figure reproduced from \citet{rodriguez06}, with permission from the authors. \label{fig:binary_SMBH}}
\end{figure*}

Binary supermassive black holes play an important role in a number of areas of astrophysics, including the formation and evolution of galaxies, galactic dynamics, and gravitational wave science. Hierarchical structure formation models predict that a significant fraction of supermassive black holes reside in binary systems \citep{volonteri03}, and these systems will have a strong impact on the central galactic environment \citep[e.\,g.][]{merritt06}. 
Simulations of binary black hole evolution in a galactic environment suggest that the inspiral efficiency (that is, the rate of decay of the binary orbital radius) may decrease at an orbital radius of $0.001~ {\rm pc} \lesssim r \lesssim 10$~pc \citep{yu02}, potentially leaving a fraction of SMBH binaries ``stalled" for extended periods of time at these orbital radii. 

The identification of a large sample of SMBH binaries would enable statistical studies of the inspiral rates in various phases of the binary evolution. This will be an important step in studies of galaxy merger rates and understanding the dynamical processes responsible for removing angular momentum from these systems, and delivering them to the gravitational wave dominated phase of evolution. The inspiral rates, and the possible existence of a ``stalling radius" are important factors in the interpretation of the gravitational wave background that will be observed by the pulsar timing array \citep{jaffe03}. Statistical studies would also allow measurements of the influence of accretion versus mergers in SMBH growth, and lead to a more precise estimate of binary merger rates \citep{burke-spolaor11}. 

Nearby binary systems that are sufficiently massive may generate gravitational radiation strong enough to enable the object to be resolved above the stochastic background \citep[e.\,g.][]{sesana09}. Whilst it may not carry a high probability \citep{burke-spolaor11b, sesana09}, the detection of both electromagnetic and gravitational wave emission from a nearby SMBH binary system would have a great scientific impact. Identification of the sky position and rough orbital solution for a nearby binary would not only raise the sensitivity of the pulsar timing array to the object manyfold \citep{jenet04}, but allow a study of the impact of the binary system on the host galaxy dynamics.

High angular resolution imaging is an effective method of searching for SMBH binaries over a wide range of orbital radii, at both high and low redshift. Binary BH candidates may be identified by surveying a large number of radio-emitting AGN (which could be initially identified in existing, lower resolution surveys) to look for dual, compact, weakly polarised, flat-spectrum radio cores \citep{burke-spolaor11}. Particular classes of AGN thought to harbour binary black holes may be targeted \citep{tingay11}. Some SMBH binary merger models predict ejected AGN, which could be revealed by astrometric measurements of AGN showing an offset from the optical host's kinetic centre. 

Source statistics are rather uncertain, since the binary black hole inspiral time-scale, and the probability that both black holes in the binary system will be radio loud, are unknown factors. At present, only one paired supermassive black hole system at a separation much less than 1 kpc is known (and supported by multi-wavelength evidence). This system, 0402+379 \citep[][]{rodriguez06}, was first identified via VLBI imaging as a candidate binary supermassive black hole (see Figure \ref{fig:binary_SMBH}). In a search of archival VLBA data aimed at SMBH binary detection, this was the only binary detected from a sample of more than 3000 radio loud AGN \citep{burke-spolaor11}. The results indicate that the VLBA is limited by (u, v)-coverage, sensitivity and dynamic range, rendering a large statistical study unfeasible. However, these crucial capabilities are realised by the SKA. The great improvement in sensitivity and dynamic range will increase the detection efficiency by allowing weaker binary companions to be identified, and weaker AGN to be searched. Improved sample selection may also significantly improve the binary detection efficiency. 

It is likely that tens of thousands of AGN must be surveyed in order to compile a significant sample of SMBH binaries, and the sensitivity of the SKA will be crucial in this regard, reducing the required integration time per source, and enabling a much larger sample of objects to be searched. Such a survey could be done in combination with a strong gravitational lens survey and \HI absorption against AGN survey. Angular resolution of $\sim$ 1 mas could resolve projected separations of 8.5 pc at all redshifts, and sub-pc separations for the nearest galaxies. This survey could feasibly be carried out at frequencies $\nu \sim 5  - 10$ GHz. 

\subsection{X-ray binary systems and \\ relativistic jets} \label{sec:X-ray_binaries}

\subsubsection{Jet Formation and Evolution}

Understanding the connection between accretion and jet production has implications for the understanding of AGN and $\gamma$-ray bursts as well as X-ray binaries. X-ray binary systems (XRBs) provide a unique tool to study the coupling between jet production and accretion flow, due to the rapid evolution of the systems through a wide range of characteristic accretion states (on the time-scale of weeks to months), and the associated rapid changes in jet characteristics \citep{fender04}. 

There exist two different classes of X-ray binary jet that show dramatically different spectral and morphological characteristics. These are the compact, steady ``hard-state" jets, and transient ``flaring-state" jets \citep[see e.\,g.][]{fender06, fender10}. The two different jet classes are associated with different characteristic X-ray states, and a typical XRB will transition between the two jet classes on varying timescales, in unison with transitions between X-ray spectral states.

Transient jets in flaring XRBs are produced during outbursts in which bright, optically thin jet components are seen moving at relativistic speeds away from the core. This type of radio jet is associated with a transition from a hard power-law X-ray spectrum to a softer power-law X-ray spectrum, and later, a thermal disk blackbody dominated X-ray spectrum. Unlike the compact, steady ``hard-state" jets, transient jets are typically resolved at VLBI scale resolution, and the optically thin, relativistic jet components are often observed to move away from the core on a timescale of hours. Relativistic ejections in X-ray binary systems can exhibit significant amplitude and structural changes over the course of a typical observation of several hours. Thus, high sensitivity, high angular resolution radio observations with good snapshot (u, v)-coverage are required to enable high time resolution ``movies" of these relativistic outflows and avoid the problems that arise from rapid evolution of the jet morphology and brightness within a single observation \citep{tingay95, mioduszewski01}. Such observations are crucial in order to tie jet ejection events to X-ray timing and spectral changes in the accretion flow. The high angular resolution component of the SKA will be of fundamental importance in this regard, particularly in the case of transient jets from neutron star XRBs, and even accreting white dwarf systems, about which very little is currently known, and which may be fainter than transient jets from black hole XRBs. It is certainly the case that the compact hard-state jets from neutron star XRBs are inherently fainter than black hole XRBs at the same X-ray luminosity \citep{migliari06}. By comparing the jets produced by accreting black hole, neutron star and white dwarf systems, the relationship between jet formation and system parameters (e.\,g.\ depth of the potential well, stellar surface, stellar magnetic field, black hole spin etc.) can be determined. Understanding the similarities and differences between disk-jet coupling in black hole, neutron star and white dwarf systems is a crucial step in understanding the jet production mechanisms and the role played by various physical parameters. 

High angular resolution will also be required to resolve X-ray binary systems in nearby galaxies from the background emission \citep{fender04}.

\subsubsection{Precise distances and luminosities}

Determining the physical characteristics of an object depends critically on knowing its distance. At present, X-ray binary distances, and therefore luminosities, have significant fractional uncertainties. Distances are typically only known to within a factor of two \citep{jonker04}. Due to the limited sensitivity of existing VLBI arrays only three X-ray binary parallax distances have been measured to date \citep{bradshaw99, miller-jones09b, reid11}. The high sensitivity and astrometric precision of the SKA will enable precise parallax distances to be measured for a large number of X-ray binary systems, and thereby enable a number of fundamental questions to be addressed, for example, by what factor can Galactic X-ray binary systems exceed their Eddington luminosities? This issue is relevant to the interpretation of ultra-luminous X-ray sources (ULXs), from which the existence of intermediate-mass black holes has been inferred. Furthermore, it is claimed that a discrepancy between the quiescent luminosities of black hole and neutron star X-ray binaries provides evidence for the existence of event horizons in black holes \citep{garcia01}. Accurate luminosity measurements are required to test this claim. Accurate distances will also enable more precise estimates of the basic physical parameters such as component masses, orbital orientation, and black hole spins, as recently demonstrated by \citet[][]{reid11} and \citet{gou11}.

\subsubsection{The formation of stellar mass black holes}

Compiling the full 3-dimensional space velocities for a large sample of X-ray binaries will provide constraints on theoretical models of stellar mass black hole formation. It is generally accepted that neutron stars receive a ``kick" during their formation, due to intrinsic asymmetries in the supernova explosion or the recoil due to the associated mass ejection \citep[see e.\,g.][and refs therein]{nordhaus10}. Supernova kicks are invoked to explain the anomalously high space velocities that are common among pulsars. It is currently not known whether all stellar mass black holes receive kicks during formation. Theoretical models predict that the highest mass black holes are formed via direct collapse of the progenitor star, with little mass ejection. These systems are not expected to show anomalously high space velocities. The less massive systems are thought to form in two stages: initially a neutron star is created in a supernova explosion, followed by fallback of ejected material which pushes the compact object over the stable mass limit, resulting in the formation of a stellar mass black hole \citep{fryer99, fryer01}. These lower mass black holes are expected to exhibit high velocities, similar to neutron stars.

Do the most massive black holes form via direct collapse, rather than a two stage process involving an initial supernova explosion? What is the mass threshold between these two black hole formation mechanisms? How does binarity and compact object mass affect the supernova explosion? Accurate distances and proper motions for a large sample of accreting black holes, when combined with data in other wavebands, can address these important questions \citep{miller-jones09a}.

Very few X-ray binaries have accurate distance \citep{bradshaw99, miller-jones09b, reid11} or proper motion measurements \citep[e.\,g.][]{mirabel01, dhawan07}. Astrometry on X-ray binaries is only possible during the so-called ``low-hard" or quiescent states. During these states, in which X-ray binaries spend the majority of their time, there exists a faint, steady, often unresolved radio jet, and there is no contribution from the bright mas-scale jet that exists during the flaring states. The closest black hole X-ray binary in quiescence, A0620-00, at a distance of 1.06 kpc \citep{cantrell10}, has a flux density of 51~$\mu$Jy/beam at 8.5~GHz and lies on the fundamental plane of black hole activity \citep{gallo06} that defines the relationship between black hole mass, X-ray luminosity and radio luminosity \citep{merloni03}. Scaling from the flux density of A0620-00 using $S_{\nu} \propto 1 / d^2$ with the best current distance estimates \citep{jonker04} for the known population of X-ray binaries \citep{remillard06}, we expect that the SKA will be capable of compiling accurate proper motions and parallaxes for several tens of black hole X-ray binaries. This would enable statistical studies of the velocities of stellar mass black holes. These conclusions are based on an assumed SKA image sensitivity of $\lesssim$ 100~nJy/beam. More robust estimates of the expected outcomes will be considered in an upcoming paper (Miller-Jones et al., in preparation).

\begin{figure}[ht!]
\centering
\includegraphics[scale=0.25]{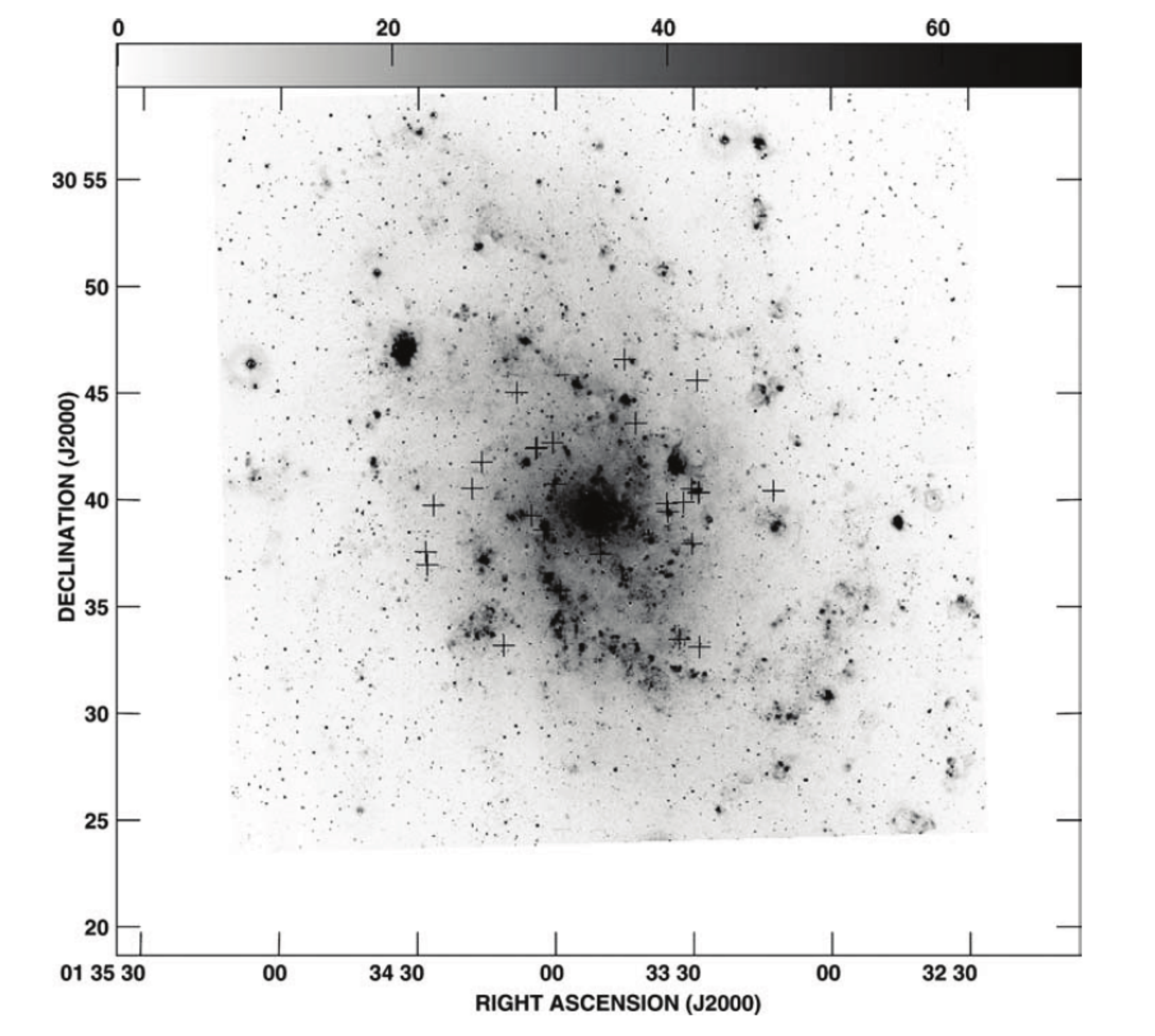}
\caption{VLA 5 GHz image overlaid on an H$\alpha$ image of the nearest relatively face-on spiral, M33. Crosses mark the locations of young, dense \HII regions. Due to limited sensitivity, the VLA only samples the top of the initial mass function --- the SKA will distinguish UC\HII regions and probe more completely the \HII region population. Figure reproduced from New Astronomy Reviews, Vol.\ 48, Hoare, ``Star formation at high angular resolution", pg.\ 1332, Copyright 2004, with permission from Elsevier. \label{fig:M33_UCHII}} 
\end{figure}

\subsection{Mapping high mass star formation in nearby galaxies}

The combined sensitivity and high angular resolution of the SKA will permit detailed studies of extragalactic \HII regions for the first time. Global aspects of massive star formation, as traced by the \HII region population, are best studied in nearby, face-on spirals such as M33 \citep{hoare04}. Figure \ref{fig:M33_UCHII} shows the location of some \HII regions in M33 discovered with the VLA overlaid on an H$\alpha$ image of the spiral galaxy. Only the top end of the initial mass function is accessible with the EVLA due to the limited sensitivity. The increased sensitivity of the SKA will be of fundamental importance in this regard. The high angular resolution component of the SKA will be able to distinguish ultra-compact \HII regions, which are young and therefore most relevant to identifying conditions at the star's birth-place. Ultra Compact \HII (UC\HII) regions are typically deeply embedded in their parent molecular cloud and so cannot be studied at optical or near-IR wavelengths.  The SKA will be capable of detecting individual UC\HII regions out to a distance of nearly 50 Mpc, and will be able to resolve UC\HII regions from their surrounding environment out to a distance of 1 Mpc \citep{johnson04}. Questions such as ``what triggers high mass star formation?" are much easier to answer in nearby spirals than in the Milky Way because a wider range of conditions can be investigated and there are no line-of-sight issues with everything lying in the Galactic plane \citep{hoare04, johnson04}.

The SKA will determine the exact location of massive star formation relative to other protostars, density enhancements in the molecular gas, shock fronts and other features of the ISM, and will enable an investigation into the relationship between properties of star formation and environmental parameters such as metallicity, pressure, turbulence, stellar density, triggering scenarios, and how star formation differs in ``burst" and quiescent modes \citep{johnson04, hoare04}.

\subsection{Other Science}

The preceding sections represent just a selection of the science case, and demonstrate that the high angular resolution component of the SKA provides a rich science case covering many areas of astrophysics, including important contributions to key SKA science. Due to length constraints, we do not consider every possible scientific application. Further science enabled by the high angular resolution component of the SKA includes \citep[see][]{godfrey11}:

\begin{itemize}
\item Strong gravitational lensing;
\item Small-scale structure and evolution in AGN Jets;
\item \HI absorption against AGN;
\item Absolute Astrometry and Geodesy;
\item Relative Astrometry: Parallax and Proper Motions;
\item Stellar winds/outflows;
\item Star-formation via studies of astrophysical masers; 
\item Stellar atmospheres:
\begin{itemize}
\item Imaging stellar atmospheres;
\item Resolving stellar radio flares;
\item Parallax and proper motions of radio detected normal stars.
\end{itemize}
\item Spatial and temporal changes in the fundamental constants;
\item Ultra High Energy Particle Astronomy at $\gtrsim$ 2 degree angular resolution via the Lunar Cherenkov technique;
\item Scattering:
\begin{itemize}
\item Probing the Intergalactic Medium via Angular Broadening;
\item Resolving AU-scale structure in the ISM via diffractive scintillation; 
\item Extreme scattering events.
\end{itemize}
\item Spacecraft tracking.
\end{itemize}

This is by no means meant as an exhaustive list of science enabled by the high angular resolution component of the SKA, and the science case will continue to develop over the coming years, as part of the Pre-construction phase Project Execution Plan or PEP \citep{schilizzi11}.  .

\section{Conclusions} \label{sec:conclusions}

High angular resolution requiring baselines greater than 1000 km provides a rich science case with projects from many areas of astrophysics, including important contributions to key SKA science. Much of the high angular resolution science can be achieved within the approximate frequency range 0.5 --- 10 GHz, or can be recast for this frequency range, and the vast majority of high angular resolution science does not require access to wide fields of view. 

\section*{Acknowledgments} 

We wish to thank the following people for helpful discussions: Joe Lazio, Clancy James, Jane Greaves, Richard Dodson, Maria Rioja and Tim Colegate.

\end{document}